        \documentstyle[epsf]{mn} 
\begin{document}
 
\title[Dust Emission from Quasars and Quasar Host Galaxies]
{Dust Emission from Quasars and Quasar Host Galaxies}

\author[Paola Andreani Alberto Franceschini Gianluigi Granato]
{Paola Andreani$^{(1)}$,
Alberto Franceschini$^{(1)}$, Gianluigi Granato$^{(2)}$\\
$^1$ Dipartimento di Astronomia di Padova, vicolo dell'Osservatorio 5,
 I-35122 Padova, Italy.\\
 e-mail: andreani@pd.astro.it, franceschini@pd.astro.it\\
$^2$ Osservatorio Astronomico, vicolo dell'Osservatorio 5,
I-35122 Padova, Italy. e-mail: granato@pd.astro.it}

\date{Accepted .....
      Received ..... in original form ...}
%\pubyear{1998}
 
\maketitle
 
\begin{abstract} 

We test emission models of circum-nuclear dust torii around quasars, 
at low and high redshifts, by using a large collection of
photometric data for an unbiased sample of 120
optically-selected objects with millimetric and sub-millimetric fluxes,
including new unpublished data. 
Under the assumption that the dust is heated by a point-like source with a
power-law primary spectrum, as defined by the observed optical-UV continuum, 
we infer the basic model parameters, such as dust masses, 
temperature distributions and torus sizes, by numerically solving the radiative
transfer equation in the dust distribution.
In addition to the substantiated statistics, an essential improvement over 
previous
analyses comes from the use of optical-UV data to constrain the primary 
illuminating continuum, which is needed to estimate dust temperatures and sizes.
Dependences of the best-fit parameters on luminosity and redshift are 
studied and the contribution of dust in the host galaxy to the observed fluxes
is briefly mentioned. This analysis constrains the properties of the
enriched interstellar medium in the galaxies hosting the quasars.
The dust abundance does not display appreciable trends as a function of 
redshift,
from $z\simeq 1$ to almost 5, and shows that dust and metals 
are at least as, and often more, abundant at these early epochs 
than they are in local galactic counterparts.  
This evolutionary pattern is remarkably at variance with respect
to what is expected for disk galaxies, like the Milky Way, slowly building 
metals during the whole Hubble time. It rather points in favour of a
much more active phase of star-formation at early epochs, probably related
to the formation of massive spheroidal galaxies.

\end{abstract} 
\vspace{5mm}

 
\section{INTRODUCTION}

The presence of dust in high-redshift objects raised 
considerable interest in the recent years for mainly two reasons.
First, there is a need to understand its effects
on the visibility of the distant universe and 
possibly to correct for the biases this implies.
Dust reddening of the light from distant
galaxies biases all optical estimators of the star-formation rate and modifies
the spectral energy distributions (SED) by mimicking an old stellar population.
It has been suggested that the early bright star-formation phase 
predicted by some models for spheroidal galaxies could have been extinguished
by dust produced by the first generations of massive stars  
(e.g. \cite{FR94}, \cite{DZ}).
Properties of the dust distribution have then to be evaluated to correct for
these various effects.

In a more positive vein, observations of dust emission
from high-redshift objects provide a vehicle of relevant information on
processes of stellar activity 
and metal yield in primeval objects, to be framed onto evolutionary models of 
galaxies.

Indeed, not only evidence is accumulating that the most important episodes of 
star-formation in the local universe occurs in dusty environments (e.g. in the 
IRAS luminous galaxies), but also
there are indications of an important role of dust in very distant objects.
This is emphasized by studies of the hyperluminous IR galaxies at high-z 
(IRAS-F10214, IRAS15307, HR10, Cloverleaf and objects detected in recent
sub-mm surveys, Hughes et al. 1997; Hughes et al. 1998b), and
by the fact that even optically ("drop-out") selected galaxies at z$\simeq$2 to 
4 (hence biased against dust obscuration) suffer dust 
extinction of 1-2 magnitudes at least in the optical at such high redshifts
(\cite{PET97}, \cite{MEU97}). Also, an integrated emission, 
in the form of an isotropic background,
and corresponding to a very energetic high redshift phase,
has been recently detected in the sub-mm (\cite{PU96}, \cite{FIX},
\cite{SCHLE}, \cite{HAUS}).

Ideal sites to look for a dust- and metal-enriched interstellar medium 
at high and very high redshifts are obviously provided by the environment of 
distant quasars, thanks to the enormous power available to illuminate and heat 
the surrounding medium. Indeed, the presence of an AGN could even mask
the presence of a forming spheroid and complicate its
investigation, but can also be exploited to identify target areas where
to look for forming structures (galaxies, galaxy clusters and groups). 

Very high metal abundances are found in clouds producing the absorption 
line systems "associated" with QSOs, showing that the close environments of 
high-z QSOs were the site of quick and
efficient star formation and metal production (Haehnelt \& Rees 1993; Hamann
\& Ferland 1993; Franceschini and Gratton 1997). 
Quantitative analysis of emission and absorption lines
would allow, in principle, to measure the metal abundances
from the equivalent
widths of atomic and ionic lines. However, these measurements are
quite complex and uncertain, and particularly in the vicinity of the quasar the
evaluation of the ionisation status is subject to systematic uncertainties.

On the contrary, the physics ruling the generation of photons by thermal dust
is much simpler and allows quite more robust and reliable evaluations of the
dust mass. Of course, this implies an extensive sampling of that portion of
the SED dominated by dust emission. Given the characteristic grain
temperatures and the typical radiation fields around quasars, this is widely 
accepted to happen also in the FIR/mm domain of radio-quiet quasars
(Sanders et al. 1989;
 Danese et al. 1992; Hughes et al., 1993, 1997; Rowan-Robinson
1995, \cite{LAW}).
\hfill\break
To this end, sensitive observations in the far-infrared to millimetric domain 
are essential.
The improvement in sensitivity of modern bolometer detectors, combined with
the favourable circumstance of the strong and negative K-correction implied 
by the steeply rising sub-{\it mm} spectra, have recently allowed to detect
and characterize with significant detail the dust properties in distant
 objects.

Many high-z sources, most of which associated with active nuclei, 
have been recently detected in the millimeter (both in the continuum and CO
line emission; see Andreani et al. 1993; Chini and Kr\"ugel 1994; Dunlop
\& Hughes 1994; Isaak et al. 1994; McMahon et al., 1994; Ivison 1995; Ohta
et al., 1996; Omont et al., 1996a and 1996b; \cite{GUI97};
\cite{I97}; \cite{HDR}).
Low-z AGN were observed by Chini et al. (1989a), Barvainis et al. (1992),
Hughes et al. (1993, 1997) and the measured spectra
nicely agree with those predicted by thermal emission from dust
components at a variety of temperatures down to $\sim$ 30 K. 

In this paper we make use of a large sample of optically selected
quasars at low and high redshifts, with far-IR data by IRAS and all
observed at {\it mm} or sub-{\it mm} wavelengths. Since both detections
and upper limits to the long-wavelength fluxes are considered, the
sample is to be considered as an unbiased optically-selected one.
\hfill\break
We interpret the observed spectra in terms 
of a geometrical model in which the dust distribution is assumed
to have simple axial symmetry, either confined to the quasar nucleus 
or covering a sizeable fraction of the hosting galaxy, and is illuminated
by the central nuclear source. Obvious evidence in support of this is
the dominance of the point-like nuclear source in the optical in both
high- and low-redshift objects.

Assuming that the radio-quiet quasars can be modeled as
a homogeneous population and are representative of the AGN class
as a whole, we build the entire infrared to {\it mm} spectrum
via observations at fixed frequencies of targets at different redshifts.

This analysis improves significantly on previous studies, as it 
exploits the whole spectral information from the optical to the {\it mm},
the optical providing in particular an essential constraint on the intensity 
of the radiation field illuminating the dust.

We adopt throughout the paper $H_0=50\ Km/s/Mpc$ and two values 
($q_0$=0.5 and 0.1) for the cosmological deceleration parameter.

\section{The data}

The sample consists of 120 optically-selected quasars with sub-{\it mm} and
{\it mm} observations and radio (either 6 or 20 {\it cm}) data. The basic 
selection
criterion is the availability of long-wavelength (in addition to the IRAS)
mm data and a quasar (rather than Seyfert galaxy) morphological 
classification.

Tables 1 and 2 list the available flux data on the whole sample:
optical magnitudes and near-IR fluxes (columns 3$\div$8),
IRAS detections or upper limits (columns 9$\div$12), sub-{\it mm} fluxes
at 450, 850 and 1250 $\mu m$ (columns 13 and 14), radio fluxes
(column 15) and absolute B magnitudes in column 16. All fluxes are in units
of mJy. Radio-loud objects are separately listed in table 2.

\subsection{Optical data}

Optical magnitudes are taken from {\cite{SSG},\cite{SGSG}, \cite{STOR}
and only for a few remaining objects from the \cite{VV} catalogue
and are corrected for intergalactic absorption according to \cite{MAD95}.
The absolute B-magnitudes are taken from the same authors and for uniformity
reported to $q_0=0.5$.
R-magnitudes are taken from the ESO/ST-ECF USNO A.1 database
(http://archive.eso.org/skycat/servers/usnoa).

\subsection{Near and Far-IR data}

NIR fluxes are taken from \cite{NEU87}, \cite{ELV}, \cite{ZIT} and \cite{TAY}.
The far-IR data are taken from the IRAS PSC and FSC, and for most of
the sources co-added fluxes were provided by IPAC based on the
SCANPI (Scan Processing and Integration Tool) program. This procedure
performs a one-dimensional coaddition of all the
IRAS survey data on the source. The sensitivity is comparable to that
achieved by the FSC (Faint Source Catalog) for point sources
(see the IPAC manual for details). 
IRAS photometry for low-z and some high-z objects have been taken
from the literature (\cite{NEU86}, \cite{SAN89}, \cite{BARV92}).

\subsection{Millimetric data}

{\it Mm} observations are partly collected from the literature
and partly obtained by the authors with the IRAM 30m at Pico Veleta. 
The low-redshift sample consists mainly of PG radio-quiet quasars
(\cite{C89a}, \cite{C89b}). High-z ones from (\cite{CK}
and \cite{I94}, \cite{I95}, \cite{MM94}, \cite{O96b}).

New unpublished data (sources 0910+56, 0953+47, 1158+46, 1247+34,
1548+46 and 2048+01)
were taken by us with the MPIfR 7-channel bolometer (Kreysa et al., 1993) at
the focus of the IRAM 30m antenna (Pico Veleta, Spain) during February
23-24 1995. 
Observations and data reductions were described several times elsewhere
(\cite{AND94}, \cite{K97}, \cite{CIM98}) and are only briefly outlined here.

For very steep inverted spectra like these,
the central effective wavelength of the system is 1.25 {\it mm}. The beam 
size is $11\ arcsec$ FWHM and the chop throw was set at $60 \arcsec $.

The average sensitivity for each channel, limited mainly by atmospheric noise,
was of 60 mJy/$\sqrt{Hz}$. The sky noise was reduced by a large factor
using the correlation among signals from the channels surrounding the
central beam.

Each source was observed with integration times in the range 4000$\div$12000 s
with the standard three beam (beam-switching + nodding) technique.
Atmospheric transmission was monitored by frequent sky-dips, which showed
zenith opacities of typically 0.2-0.4.
Calibration was performed using Uranus as primary calibrator and quasars of 
known
flux as secondary calibrators. The pointing accuracy was checked each hour and 
the
average error turned out to be $ 3 \arcsec $.

Data were reduced assuming that the target sources are point-like, or at least 
that
their extent at {\it mm} wavelengths is smaller than the size of the
central beam. The other six channels were then exploited
to reduced the sky noise in the central one, by subtracting from the signal
in the central channel the average value of the atmospheric level computed
over the outer six ones, for any given elementary integration.
%Note that this procedure only gets rid of that part of the sky fluctuations
%with correlation length smaller than the receiver array ($60 \arcsec $),
%i.e. for motions at {\it short} wavelengths.
%However, note that the dominant part of the atmospheric noise is produced
%by motions at {\it long} wavelengths (Penzias and Burrs, 1973; James
%and James, 1989), while high frequencies (5$\div$20 Hz)
%fluctuations do not contribute much to the noise (Ade et al., 1984) and
%are smoothed out by the fast wobbling of the secondary.
%According to Andreani et al. (1990), the correlation length for
%convective shells at these wavelengths is of the order of several
%tens of centimeters, i.e. only fluctuations generated at altitude larger
%than 2000 m above the telescope survive to the double-switching subtraction
%and contribute mainly to the noise. At altitude higher than 5000m, however,
%the residual water vapour is very low and the residual noise is also
%low.

\subsection{Radio data}

Radio VLA data were used to divide the two quasar populations 
on the basis of the ratio $R$ between the radio flux density at 5 GHz,
 $S_{5GHz}$, and the optical flux at 4400 \AA ~, $F_B$,
the radio-quiet objects having  $R = \frac{S_{5GHz}}{F_B} < 10$
(e.g. \cite{KEL}, \cite{FAL96}).
Data are taken from \cite{SSG}, \cite{SGSG}
and from \cite{KUK}, \cite{VV}.

\section{EMISSION MECHANISMS IN RADIO-QUIET AGN}

Though there has been an active discussion during the past several years about 
the
origin of the AGN continuum at infrared wavelengths (the two alternative 
interpretations being either a synchrotron radiation from the AGN itself, or 
emission by circum-nuclear dust), the more recent tendency was to clearly favour
the latter, at least for radio-quiet objects.
Crucial observations in this sense were the comparison of the millimetric fluxes
with IRAS survey data, showing sub-millimeter slopes of the SED with spectral
indices in excess of 3.

In principle, synchrotron self-absorbed (SSA) sources can have a spectral-index
steeper than 2.5 if the electron energy distribution is dominated by a
steep low-energy electron population (de Kool \& Begelman 1989;
Schlickheiser et al. 1990). In practise, however,
with a non-thermal turnover in the FIR (at 100-200$\mu $m), such steep
indices are not reached until mm-wavelengths and hence the available sub-mm
data clearly indicate thermal processes as the origin of the IR flux
in AGN.

Indeed, dust organized in a toroidal structure surrounding the central engine,
as envisaged by unifying models of the AGN activity, 
and with a wide temperature distribution up to $\sim$ 1500 K (where grains
begin to sublimate), re-radiates thermally at $\lambda \geq 1\mu m$ photons 
absorbed
from the primary source (Neugebauer et al. 1987, \cite{SAN89}).
 This interpretation in  terms of dust emission 
naturally accounts for the basic observed features of the long-wavelength
spectrum of AGN. In particular, it explains the very steep spectral indices
for AGN in the sub-mm as grey-body emission of thermal dust with absorption
coefficients depending on frequency as $\alpha(\nu) \propto \nu^{1-2}$. It also
nicely explains the minimum in the SED often observed in AGN 
at $\lambda \sim 1\ \mu m$ and due, in this framework, to the cross-over
of the primary or stellar continuum with the emission spectrum from dust
at the sublimation temperature (e.g. \cite{SAN89}, Granato \& Danese 1994).
Thermal emission at these sub-mm wavelengths is also supported by the
detections of CO emission from low and high-z radio-quiet quasars
(see the review by \cite{BAR97}; \cite{SET}).

A more complex situation is considered by \cite{RR95}, who interprets
the infrared spectrum of radio-quiet quasars (RRQs) as contributed both by
an AGN component with peak emission at $\lambda = 3-30 \mu m$
 and a starburst emission at $\lambda > 30 \mu m$.

At wavelengths longer than 100 $\mu m$ in the object's rest-frame, it is 
possible
that thermal emission from the host galaxy dominates the FIR/{\it mm} SED
of the quasar (\cite{DAN}, \cite{FAD}). Testing this is impossible with the 
poor spatial resolution of current instrumentation (which would rather require the 
resolving power of large millimeter arrays in plan).  

%At longer wavelengths interpretation is more difficult because of the
%small amount of data. Furthermore, the mm emission of low-z objects could be
%underestimated if emission is extended. For nearby objects the beam width
%of single-beam experiments only covers the very central region 
%and this emission could be not negligible in cases in which
%several dust components at temperatures down to $\sim$15 K are present
%in the host galaxy.

\subsection{A model of dust emission in quasars}

We then assume in the following that the observed FIR/{\it mm} emission 
is due to dust surrounding the AGN.
We further make the rather coarse assumption that the optical properties
of this dust are roughly similar for all sources,
with a standard Galactic composition.

The dust illuminated by the central nuclear source reaches an equilibrium 
temperature which is a function of the intensity of the radiation field (hence 
of
the distance from the central source) and of the dimensions and chemical 
composition of the assumed grain components. We model the SEDs of radio-weak 
quasars
following \cite{GD94}. The observational data are fitted with model
spectra produced by a numerical code which solves the radiative transfer 
equation 
in a circumnuclear dust distribution.

The latter step is required since in the torii predicted by unified models the
dust emission is self--absorbed even in the near-- and mid--IR.  The
code is quite flexible in dealing with different geometries and composition of 
the
dusty medium, the only restriction being axial symmetry. It thus allows
a wide exploration of the parameter space.

The inner radius $r_{in}$ of the dust distribution is set by the grain
sublimation condition, at $T_s=1500$ for graphite and $T_s=1000$ for
silicates. This translates into the condition 
$r_{in}\sim 0.5 \sqrt{L_{46}}$ pc, where
$L_{46}$ is the luminosity of the primary optical--UV emission in units
of $10^{46}$ erg $s^{-1}$.

The details of the model, as well as the effect of the various free
parameters, have been widely discussed by \cite{GD94} and
\cite{GDF}. Here we focus on
the simplest geometry which is in reasonable agreement with the available
observations, the ``flared disc'', in which the scale height of the
torus along the z-axis increases linearly with the radial distance

\begin{equation}
\rho(r,\Theta) = C r ^{-\beta} exp(-\gamma | \cos \Theta |)
\end{equation}

\noindent
where $ \Theta _h \leq \Theta \leq \pi - \Theta _h $ and $\Theta _h > 0 $
 is the half opening angle of the dust--free cones and 
$\gamma = 0$ for a torus-like distribution.
The dust distribution within the disc is taken homogeneous
($\beta = 0$). Model parameters
are then {\it (i)} the outer radius $r_{max}$ of the dust
distribution; {\it (ii)} the absorption $A_V$ along typical obscured
directions, or the corresponding optical depth $\tau$ at 0.3 $\mu m$,
where $A_V \simeq 0.61 \tau$; 
{\it (iii)} the angle $\Theta$ between the torus axis and the line-of-sight
or equivalently the covering factor $f=\cos(\Theta_h)$ of 
the torus; {\it (iv)} the value of $\gamma$ in eq.(1).
For most of the sources free parameters in the fit are only
$ r_{max}$ and $A_V$,   while $\gamma$ was set to 
0 and the covering factor has been fixed to $f=0.7$, a value not 
inconsistent with the requirements of unified models of AGN (Granato et
al. 1997) (see table 3).
Only two objects (0759+651, 1334+243)  were fitted with a dust distribution
completely covering the nucleus, thicker at the equator than at the poles
(in this case $\gamma > 0$ and $\Theta _h = 0$).

To the first order, the outer radius $r_{max}$ mostly determines the broadness 
in wavelength of the 
IR bump arising from dust re-processing, whilst the second parameter, $A_V$, 
controls
the near-IR slope of the SEDs as observed from obscured
directions, as well as its anisotropy.  Therefore $A_V$ is in principle
testable from measurements of the SEDs of obscured AGN or from the anisotropy 
of mid--IR emission. The covering factor is obviously related to the
relative power reprocessed by the dust in the IR bump
and the primary optical continuum flux.

In the simpler geometry with $\gamma = 0$
the total dust mass is given by the product
between the assumed disc volume $V=\frac{4}{3} \pi  \cos \Theta_h$ and the
(assumed constant) dust density $\rho=\frac{\tau_\nu}{k_\nu (r_{max}-r_{in})}$,
where $k_\nu$ is the absorption coefficient per unit dust mass
and $\tau _\nu$ the equatorial optical thickness.
This gives:

\begin{equation}
\frac{M_{\rm dust}}{M_\odot} \sim 0.2 \, ({r_{max} \over r_{in} })^2
({L \over L_{46}}) \cos\Theta_h \, A_V
\end{equation}

\noindent
as the total mass in dust as a function of the basic model parameters.

\section{RESULTS}

\subsection{Masses and sizes of the dust distributions}

Figure 1 shows the spectral energy distribution as $\nu L_\nu$ versus $\nu$
in the object's rest-frame of all sources in the sample with either FIR or 
sub-{\it mm} detections. The spectra are normalized at the 100 $\mu m$ datum.
Some information on the source and best-fitting parameters is reported in Table 
3.
Significant constraints on dust masses and sizes are estimated for sources 
with only upper limits to the long-wavelength fluxes.
The thick solid curves in Fig.1 are the best-fit SEDs, while the dotted line is 
the broad-band optical-{\it mm} spectrum of a typical local spiral galaxy, used 
for 
comparison. We adopted to this purpose the SED of M51, with its proper 
normalization corresponding to an assumed distance of 9.6 Mpc
(see Silva et al. 1998).

Clearly, while for local objects the hosting galaxy can 
contribute to some extent at the longest wavelengths, for high-z quasars 
the observed spectra are far in excess of what would be expected
from a redshifted SED of a normal galaxy. This reflects the
much increased power of the illuminating source with respect to what happens
in local spiral galaxies.

The overall quality of the fits is remarkably good, if we consider 
that essentially two parameters ($r_{max}$ and $A_V$, the inclination angle 
$\Theta$
not affecting significantly the spectrum for most of the sources) 
are free to model the spectral shape.

Figure 2 plots the mass $M_{d}$ of the circum-nuclear dust distribution
derived  from our spectral best-fits, as a function of 
the B-band luminosity, L$_B$. The latter measures the energy output of the 
primary
nuclear component.  Although there is a tendency for the lowest-luminosity
objects to have lower values of $M_{d}$, the correlation is generally very poor 
and 
the scatter very large, in particular there is no dependence on $L_B$ for 
$L_B>3\ 10^{11}\ L_{\odot}$.
This is all but reminiscent of a simple scaling of the mass in dust with the 
quasar luminosity.

Figure 3 shows the distribution of the dust mass M$_d$ as a function of 
redshift.
The down-pointing arrows correspond to objects with only upper limits to the 
long
wavelength fluxes. The solid horizontal line marks  
the typical dust mass of a nearby spiral (Andreani \& Franceschini 1996).
The lower panel shows the dust masses computed in a $q_0 =0.5$, while
the upper panel reports the corresponding estimates in a low density
universe ($q_0 =0.1$).
Low-z objects present a large spread in the values of M$_d$, from $10^4$
to $10^8$ M$_\odot$. The majority of high mass objects are at redshift larger
than 1. An obvious bias selects at high redshift only luminous and massive
objects and the lower right part of the plot in figure 3 is uncovered.
 
Figure 4 plots the best-fit outer radius $r_{max}$ of the dust distribution 
versus redshift. $M_d$ and $r_{max}$ are very tightly related in our 
model, since most of the mass comes from the outer, less illuminated,
lower temperature regions of the dust distribution. Both of these 
parameters are mainly constrained by the long wavelength ({\it mm} or 
sub-{\it mm}) flux data.
Then the plots in Figs. 3 and 4 display quite a similar behaviour, as expected.

In both cases the low-z objects show widely dispersed values 
in the parameters ($M_d\sim 10^4 - 10^8\ M_\odot$) and  $r_{max}$
from 200 pc to less than 3 kpc.
For the high-z subsample, typical values for $M_d$ and $r_{max}$ are quite
significantly larger, with $M_d$ in excess of $10^7$ and $r_{max}$ in excess 
of 1 kpc. 

Note that the procedure of the dust mass estimation is
quite robust: we believe that uncertainties in the model
parameters affect its value by less than a factor two, and our basic
conclusions are rather secure.

\subsection{A far-IR to optical color-magnitude diagram for quasars}

We report in Figure 5 the ratio of the far-IR ($L_{FIR}$) to the B-band 
($L_{opt}$) 
luminosities as a function of $L_{bol}$ (in solar units).  
$L_{FIR}$ is obtained by integrating the rest-frame SED from 1$\mu m$ to
1 mm, while $L_{opt}$ is found by integrating the spectrum between 1200 and 9000 
\AA~.
 It should be first noticed in Fig. 5 the wide range covered by the bolometric 
luminosity,
which reflects the enormous range of nuclear power.

Figure 5 shows that there is a rough equipartition of the light directly 
escaping
from the nuclear source -- and measured by the optical flux -- and that
absorbed by dust and reprocessed in the infrared. 
A remarkable effect apparent in the figure is that,
while showing a large spread in the bolometric luminosity, our sample objects
cover a very narrow interval in the infrared to optical luminosity ratio 
$\frac{L_{FIR}}{L_{opt}}$ between $\sim -0.1$ and $\sim 0.2$.
Only two objects do not follow this rule: 0759+651 presents a
distinct behaviour with
$log(L_{bol})=13.4$ and $log(\frac{L_{FIR}}{L_{opt}})=1.06$
due to extinction of the optical light.
Note, however, that this object is the only one of the sample
 which was not optically
selected but it was discovered by IRAS.
 0844+349 on the contrary (at $log(L_{bol})=12.6$ and  $log(\frac{L_{FIR}}{L_{opt}}) \sim -0.36$) presents
a high optical luminosity with respect to the FIR one
(see the corresponding spectrum in Fig. 1).

In the unified AGN scheme the color axis, $\frac{L_{FIR}}{L_{opt}}$,
is related to the viewing angle of
the torus, with more inclined objects lying in the upper part of the
diagram, because of the larger FIR emission. The almost complete
lack of objects in this region of the diagram emphasizes that optical surveys, 
like the present one, provide a strongly biased census of the quasar
population: they almost entirely miss extinguished objects.
Our model of circumnuclear dust 
(whose correctness is proven by the good match with the observed spectra)
predicts that an unbiased sampling of the quasar population (e.g. a selection
 at long wavelength where $\tau\sim 0$) would found roughly 25\% of the objects
to be
 highly extinguished quasars (i.e. with $\Theta _h \geq 70^\circ$) .

The origin of the small scatter in the far-infrared to optical-UV luminosity
ratio of the optically selected quasars is not clear and has to
be further investigated.
In particular, there is no evidence for a population of quasars characterized
by very low far-IR emission, i.e., in the present interpretation, by negligible 
amounts of circum-nuclear dust. Indeed, the distribution of upper limits in Fig. 
5
may originate from the same parent population of the long-wavelength detections.
The lack of objects with low IR/optical luminosity ratio may be simply due
to the sensitivity limits of the present millimeter observations but it may
also be indicative of a narrow intrinsic distribution in the FIR to optical
luminosity ratios.
Extensive surveys with sub-mm telescopes (SCUBA, CSO) are needed to solve
this ambiguity.

\subsection{Radio-loudness versus radio-quiescence}

Figure 6 compares the radio power to the quasar optical luminosity. Filled 
squares correspond to the radio-loud sub-population, the other symbols to
the radio-quiet one.
The radio-loud objects are selected according
to their ratio $R = \frac{F_{radio}}{F_{opt}} > 10$.

The radio-quiet quasars show a fairly well defined (not far from linear) 
correlation between the two luminosities, in spite of the large fraction of
radio upper limits, suggestive of a common physical mechanism originating
the two emissions.

The radio activity shows up as a dramatic increase, by several orders of 
magnitude,
of the radio power occurring in a fraction of the highest optical luminosity
quasars. A very similar plot of optical versus radio luminosities, scaled down 
in 
luminosity, was found by Calvani, Fasano, \& Franceschini (1989) for radio
galaxies.
Low-luminosity AGN seem in any case to provide an unfavourable environment
for the ignition of radio activity.

To understand if radio activity is related to some unusual values of the IR to 
optical
luminosity ratios, we have reported in Figure 7 the far-IR to optical versus
far-IR to radio luminosity ratios. The radio active quasars show the lowest
$\frac{L_{100 \mu}}{L_{radio}}$ values. In fact, radio activity does not entail
any unusual effects in the $\frac{L_{100 \mu}}{L_{opt}}$ flux ratio and seems
to be independent of the optical and FIR properties.

The origin of the millimetric flux in radio-loud objects is finally studied
in Figure 8, a plot of $\frac{L_{1 mm}}{L_{radio}}$ versus
$\frac{L_{1 mm}}{L_{opt}}$. Only very few radio-loud quasars have 
$\frac{L_{1 mm}}{L_{radio}}$ values consistent with a power-law synchrotron
extending from the radio centimetric to the mm. In the vast majority of
objects the mm emission is in excess of the radio one, indicating that
at wavelengths shorter than 1 mm either a high frequency synchrotron 
self-absorbed,
or more likely dust emission is overwhelming the non-thermal one.

\section{DISCUSSION}

Evidence in favour of the idea that dust emission processes dominate
the infrared through millimetric emission spectrum of AGN has steadily
increased in the last years.

We are able to confirm this indication with the present optically 
selected sample, but mostly limited to low-redshift quasars with far-IR IRAS 
detections and upper limits to the millimetric fluxes. 
In more than 70\% of the sources
the sensitivity of the mm observations combined with the IRAS 
detections
(see spectra in Figure 1) are enough to exclude a synchrotron component based
 on the steepness of the continuum.

Unfortunately, at redshift larger than 1, with the exception of
three quasars, objects with mm or sub-mm detections (favoured
by the strong K-correction) were not usually detected by IRAS and a similar
comparison with mm and sub-mm fluxes was not typically possible.
The three exceptions are QSO0838+3555 (at z=1.77), QSO1634+706 (at z=1.34)
and the well studied QSO1413+117 (at z=2.546). In these cases,
with the addition of BR1202-725 (at z=4.69) detected at 150 and 200 $\mu m$
(rest frame), a dust emission spectrum with its very steep
convergence in the RJ low-frequency limit provides an extremely good fit the
the observed SED.

Additional support for this interpretation may also be found
if we consider the ubiquitous presence of a dip in the SEDs of Figure 1
 around $\lambda = 1\ \mu m$ for all quasars with good near-IR data
(naturally explained by the dust sublimation hypothesis),
and from the generally good agreement provided by the dust re-radiation model
with the broad-band data.

With this working hypothesis in mind, we have exploited
long-wavelength observations of quasars to investigate properties of
the dusty interstellar media in local and up to very distant objects. 
Original motivation for this was to achieve a new independent information 
channel, whose results about the metal enrichment in extremely distant objects
are to be compared with independent analyses of emission and absorption metal
lines.

The key to achieve this was to combine long-wavelength observations (both IRAS 
and mm/sub-mm fluxes and upper limits) of thermal dust
emission with optical-UV data on the primary illuminating spectrum. This was
essential to constrain the two fundamental parameters of the dust model, i.e.
the size of the dust distribution (or the average distance of dust from the 
central illuminating source) and the equilibrium temperature of dust grains.
The obvious remaining problem was due to the poor description of
 the quasar SED allowed by the available data, which imply somewhat
degenerate sets of solutions in terms of these two fundamental parameters,
 of dust grain compositions, and so on.

We emphasize, however,
 that exploitation of all available data allows quite a robust and
unbiased evaluation of the dust mass $M_d$, which should be independent of the
quasar luminosity and redshift.

Given the relative simplicity of the physics ruling dust emission,
the hope was to obtain meaningful information from a limited dataset.
This section is devoted to discuss our findings of the previous section,
starting with a preliminary discussion about the origin of the quasar's long 
wavelength flux.

\subsection{Intervening galaxies: quasars with DLy$_\alpha$ systems}

Let us check first the possibility that the observed {\rm mm} emission in the
highest redshift quasars might be due to galaxies intervening along the line
of sight.
Recent deep surveys with the SCUBA bolometer array on JCMT (Hughes et al. 1998a;
Smail et al. 1997; Barger et al. 1998) have revealed that the millimetric sky is
 populated at faint ($\sim mJy$) flux limits, such that millimetric
surveys are confusion limited already at a few mJy at 850 $\mu m$.
This indicates that there is a non-negligible chance of {\rm mm} detection in a random
direction at these fluxes. A way to check for the presence of gas-rich galaxies
along the quasar sight-line, producing a millimetric signal unrelated to the
quasar, is offered by absorption line studies in the quasar continuum, in
particular looking for damped Ly$_\alpha$ systems which are believed to 
correspond to already assembled galaxies.

High resolution spectra available for some objects in our sample show indeed the
presence of such line systems. 
We consider here the following four: $z_{abs}=4.3829$ in the line of sight
of BRI 1202-0725, $z_{abs}=1.7764$ in the line of sight of QSO1331+1704,
$z_{abs}=4.08$ in QSO2237-0607, and $z_{abs}=3.39$ in the line of sight of 
QSO0000-26 (\cite{LU}, \cite{MOL}).
%and $z_{abs}=3.6617$ in the line of sight of 2212-1626

We have used the metal abundances of the absorber estimated from the absorption
lines as an estimate of the gas metallicity. The amount of the corresponding 
dust mass is then inferred and compared with the one required to fit
the FIR/sub-{\it mm} emission. This implies the assumption
that the metallicity of a system is related to its dust content, a relationship
observed in the Galaxy and in nearby objects
(e.g. Issa et al. 1990; Sodroski et al., 1995; \cite{AF96}). For high
redshift objects, Pei et al. (1991) computed
the ratio $k$ between the B-band optical depth, $\tau_B$,
and the hydrogen column density, $N_H$, (with respect to the Galactic value
$k_G$), which gives an estimate of the dust-to-gas mass ratio, as a
function of the metallicity. We estimate in this way the parameters
$k/k_G$,  $\tau_B$ ($ k ~ N_H (cm^{-2})/10^{21}$) and $M_{dust}$ (see values
in Table 4).
% $k/k_G=0.01$, $k=8~10^{-3}$. \tau_B = 3.2 ~10^{-3}$.
%From the observed metallicity of the galaxy in the line of sight of
%BRI 1202-0725 ([Fe/H]=-2.2, Lu et al., 1996),
%This means that for $N_H = 4 ~10^{20} cm^{-2}$, the
%B-band optical depth, $\tau_B = k ~ N_H (cm^{-2})/10^{21} = 3.2 ~10^{-3}$.
%this latter depends on the dust extinction efficiency 
%$Q_{ext}$ and average dimension $a$:
%
%$$\tau_B = N_{gr/H atom} ~ \pi a^2 Q_{ext} N_H $$
%
$M_{dust}$ can be estimated from the $k$ value in the framework
of a dust model. Using the grain model mentioned in \S 3.1
we can infer the mass in g per H-atom, $\Re_{sys}$, from:

$$ \Re_{sys} = \Re_{mod} ~ k_{obs}/k_{mod} =
N_{gr/H atom} ~ {4 \over 3} \pi a^3 \rho_{gr} $$

%= 1.47 ~10^{-26} $$

\par\noindent
where $N_{gr/H atom}$ is the number of grain per H-atom.
The estimated dust mass becomes:

$$ M_d = N_H ~ A ~ \Re _{sys}  $$

\noindent
where $A$ is the projected area
of the Galaxy. The dust masses, listed in column 8 of Table 4,
turn out to be very small even if the dust is spread over a diameter of
10 kpc and would produce a negligible {\rm mm} emission.

These results are in agreement with the analysis by Franceschini and Gratton 
(1997).
Absorbing systems in the quasar
line of sight with escape velocities larger than 5000 km/s are not
associated with the quasar, and have typical metallicities much less than solar
(in the range Z/Z$_\odot \sim $ 0.1$\div$0.001),
in marked contrast with the {\it associated} systems ($v < 5000\ km/s$) 
which have metallicities Z/Z$_\odot \simeq 1$ or higher.
Note that only for a solar metallicity ($k/k_G \sim 1$) the amount of dust mass 
in the absorber would be of the order of $10^{7} M_\odot$, close to that in
present-day spiral disks, but still significantly less than that inferred from
the far-IR flux.

Altogether, the observed {\rm mm} emission is very likely due to dust
{\it associated} with the quasar itself.

\subsection{Dust in the quasar and in the host galaxy}

Some limitations of the present analysis have to be considered
before drawing conclusions.

The first is that
the observational SEDs are interpreted assuming that most of the IR-{\it mm} 
flux comes from dust illuminated by the AGN itself and that the dust in 
the flared disc is homogeneously distributed from the inner sublimation 
surface up to kpc-scale distances.

For most of the local objects only upper limits
to the {\rm mm} fluxes are available. So, it cannot be excluded that for some 
of these we are missing some contributions 
of cold dust from the hosting galaxy at $\lambda \geq 100\mu m$ 
(most of the {\rm mm} observations were collected with large
antennas fed with single channel bolometers, having therefore a field of view
limited to a few arc-sec). Indeed, Danese et al. (1992) and \cite{RR95} 
already suggested that the FIR spectrum at $\lambda \geq 60 \mu m$ in 
low-luminosity AGN may be contributed by cold dust in the surrounding galaxy 
located at large radial distances from the AGN and with            
dust masses larger than 1$\div$3 $\ 10^7$ M$_\odot$.  The relative contribution
at far-infrared wavelengths of the nuclear and extended component, and hence
also the total dust mass,
could only be settled by high resolution mapping with planned large
{\rm mm} arrays (e.g. LSA/MMA).

The SEDs of high-z quasars are often poorly constrained, due to the fact 
that IRAS was not sensitive enough to detect them
(a few exceptions appear in Fig. 1).
% In these cases the
%sub-{\it mm} data are the only information we have on the circum-nuclear
%interstellar medium, IRAS providing only a constraint on the dust
%temperature distribution.
In any case, comparison of our model with the data suggests the existence of a 
large 
gaseous structure around the nucleus (with 1-5 kpc radius), including a dust 
mass 
of up to 10$^9$ M$_\odot$, and a corresponding gas masses of typically 
$M_g = 10^{10}$ to $10^{11}$ M$_\odot$. 

The mass of the dust distribution around quasars shows a very weak dependence
on the quasar optical luminosity $L_B$ (Fig. 2), and consequently on redshift
(Fig. 3). Quasars with the lowest $L_B$ tend to
have less circumnuclear dust, but with a large scatter.
%In principle, we would expect that the mass $M_d$ and the size $r_{max}$ of
%the dust distribution should properly account
%for the power of the primary continuum and its evolution with cosmic time.
%
%In practice an obvious bias operates in Figure 3: the more luminous AGN
%found at the higher z are also likely to be hosted by higher mass 
%galaxies.
%Small mass objects at high-z escape detection and the lower right part of the
%plot cannot be sampled by present instruments. 
%
%What is clear is that not many high mass objects are found in the nearby
%universe. 
%
Indeed, the dependence of the circum-nuclear dust mass on redshift
does not seem to follow the time evolution of the quasar luminosity function
(though it is significantly affected by the Malmquist bias).
This is also illustrated by Fig. 2, where the dust mass shows a very weak, 
certainly not linear, dependence on nuclear luminosity.

Concerning the overall extent $r_{max}$ of the dust distribution, Fig. 4
reveals that local quasars and Seyfert galaxies (at $z<0.2$) show dust
structures out to a maximum radius of 1 kpc. For higher redshift quasars
$r_{max}$ increases up to values of several kpc, indicating that these
structures are not merely circum-nuclear dust torii, as seen in local Seyfert
galaxies, but have rather scale-lengths comparable or even larger than those
of the bulge of the putative host galaxy. How reliable are these estimates? We
claim that whenever the IRAS flux limits, combined with {\rm mm} detections,
meaningfully constrain the model SED, there is no margin to decrease $r_{max}$
by increasing the dust temperature $T_d$ and reducing the mass $M_d$, because
this would imply a strong and unacceptable violation of the IRAS limit at
typically 60 to 25 $\mu m$. Certainly in all these cases our estimate of
$r_{max}$ should be quite reliable.

{\it The result about $r_{max}$ is indicative of a medium heavily enriched by 
dust and 
metals on a quite
large scale around the central AGN, as would be expected for an already large 
and massive galaxy in an advanced stage of chemical evolution}.
%These observations then probe the ISM around the quasar on a much larger scale
%than allowed by analyses of emission lines in the quasar spectrum.

At the sensitivity limits of the present millimetric observations there is no
evidence for a sub-population of quasars with no or significantly depressed
far-infrared emission. As illustrated by Fig. 5, the far-infrared emission
scales typically linearly with the optical quasar luminosity.
{\it Dust seems to be a ubiquitous presence at any redshifts, whenever a
sensitive enough observation has targeted it}.

Though needing confirmation by further sensitive observations with the 
new-generation
sub-millimetric imagers (SCUBA, CSO), this result, also in combination with
metal line observations, tends to provide support to the view that
the quasar phenomenon follows the formation of a large spheroidal galaxy
(e.g. \cite{HF}, Franceschini \& Gratton, 1997), rather than 
triggering and preceding it (Silk \& Rees, 1998).

The time evolution seen in Figs. 3 and 4  does not seem easily interpretable 
in terms of a unique class of objects. Where all of the observed dust mass at
high z has gone? Was it swept away from the quasar? Was it consumed to form
stars in a late phase of star formation? (\cite{EE}).
Note that the difference between low and high redshift objects would be
further enhanced in an open universe: dust masses at redshifts larger
than 2 for $\Omega \sim 0.01$ are higher by a factor between 2 and 5.

The alternative possibility is that we are observing at high- and 
low-redshifts two distinct quasars populations.
Environmental conditions at high redshifts could favour the formation 
of higher mass black holes, while low-z AGN may be related to the refueling 
of small dead black holes in later type galaxies, where gas is still available
(Haehnelt \& Rees 1993; Cavaliere \& Vittorini 1998).

%Figures 3 and 4 also indicate that the properties of the dust distributions
%around quasars kept the same over a very large redshift interval from $z=5$
%to less than 1. A similarly large range of ages and formation epochs has been
%recently found to characterize early-type galaxies in the field 
%(Franceschini et al. 1998), which current observations of
%massive nuclear black-holes associate with the past quasar activity.

We finally note that our optically-selected sample has a clear bias
against optically extinguished quasars with the torus viewed edge-on,
as we have shown in Sect. 4.2.
In addition, an entire early phase in which the object is still obscured by the 
products of massive galaxy-forming starbursts could have escaped detection.
Only large-area surveys at long-wavelengths, such as those to be performed 
with the ISO, SIRTF and FIRST missions, will be able to discover them.

\section{Summary and Conclusions}

We have performed a detailed analysis of broad-band spectra for a rich 
unbiased sample of optically-selected quasars with far-IR and {\rm mm}
data, with an emphasis on the portion of the SED dominated by dust thermal
emission.                                                                
The exploitation of flux data over a wide frequency interval has allowed us 
to successfully constrain the basic parameters of the dust distribution 
around the quasar, i.e. the size $r_{max}$, temperature $T_d$ and total 
dust mass $M_d$.      
\hfill\break                                                          
Our main results are summarized below.                                   
            
\begin{itemize}                                                              
\item                                                                         
We find that, when detailed enough data exist, a model of emission by 
circum-nuclear dust illuminated by the optical-UV flux of the quasar 
provides a very good fit to the observed SED, including the steep 
sub-millimetric 
convergence and the observed minimum at around $\lambda = 
1 \ \mu m$. This supports the concept of a thermal origin for the IR SED 
in radio quiet quasars and tends to rule out alternative interpretations. 
\hfill\break
Radio activity in quasars does not seem to entail unusual properties of far-IR 
and optical emission with respect to the radio-quiet population. 
  
\item 
The mass of dust around high redshift ($z>1$) quasars is typically observed 
to fall 
between $10^7$ and $10^9$ M$_\odot$, with no appreciable dependence on $z$
over a very 
large interval from $z=1$ to $z=5$. Similarly, the radial size of the dust
 distribution $r_{max}$ runs from $r_{max} \simeq 1$ kpc to $\simeq 10$ kpc over
the same  redshift interval. 
These results indicate the presence around quasars of massive dust 
distributions on scales comparable with those of the hosting galaxy. 
The latter have to be already large systems observed in advanced stages 
of chemical evolution during or after a very active phase of star and metal 
production. 
This evolutionary pattern looks fundamentally different from that expected 
for disk-dominated galaxies, and requires an early accelerated phase 
of star-formation, probably related to the formation of a massive spheroid.

\item 
At the sensitivity limits of the present data, the ratios of the bolometric 
far-infrared to optical-UV luminosities show a narrow distribution, 
implying a roughly constant fraction of optical-UV light being reprocessed 
into the IR, with no evidence for a population of metal-poor optically 
dominated objects. 
Although more sensitive {\rm mm} surveys (e.g. with SCUBA) are needed to
 better probe the 
faint end of the distribution, this tends to support the view that the 
quasar phase parallels and/or follows the formation of a large hosting galaxy, 
rather than preceding it. 

\end{itemize}

\section*{ACKNOWLEDGEMENTS}
We would like to thank S. Cristiani for critically reading this work,
L. Danese for instructive discussions and F. La Franca for helpful
comments.
Part of the data used in this work were taken with the SCANPI procedure,
developed by the NASA Archival center for IRAS Satellite (IPAC) operating
by JPL.

This work was partially supported by ASI (Italian Space Agency).

\newpage
{\bf Figure captions}

{\bf Figure 1} Spectral energy distributions (S.E.D.s)
 of the quasars of the sample.
Each panel reports the name and the redshift of the object. The solid
line represents the predicted S.E.D. according to the model described in
the text (\S 3). The dashed-line is the SED of a template nearby dusty spiral
with a dust mass of $10^7 M_\odot$.

{\bf Figure 2} The dust mass computed according to the model prescriptions
described in the text (\S 3.1) as a function of the B-band luminosity.
Down arrows correspond to those sources with only upper limits at
FIR/sub-mm wavelengths.

{\bf Figure 3}
The dust mass as a function of redshift in an open universe
($q_0=0.1$, $\Lambda=0$) (upper panel) and in a flat universe
($q_0=0.5$, $\Lambda=0$) (lower panel). Down arrows correspond
to those sources with only upper limits at
FIR/sub-mm wavelengths. Note how the difference between the nearby and the
distant source masses increases for an open universe.

{\bf Figure 4} The maximum radial size, $r_{max}$, of the torus surrounding the central
source for the quasars of our sample.

{\bf Figure 5} A colour-luminosity $log(\frac{L_{FIR}}{L_{opt}})$ - $L_{bol}$
plot for the quasars of the sample. Filled
circles correspond to low-$z$ ($z \leq 1$) objects. Open squares to intermediate
redshift ($ 1 < z \leq 2$) sources, asterisks to high$z$ ($z \geq 2$) ones.
The two outliers are 0759+651 (at $L_{FIR}/L_{opt} = 1.06$)
and 0844+349 (at $log(\frac{L_{FIR}}{L_{opt}}) \sim -0.36$).

{\bf Figure 6} Radio power as a function of the B-band absolute magnitude. Filled squares
correspond to radio-loud objects, while open circles to radio-weak sources.
Down arrows show those objects observed at 5 or 1.4 GHz but not detected,
hence assumeed to belong to the radio-weak population.

{\bf Figure 7} A colour-colour plot showing the behaviour of the rest-frame
luminosity ratio, $L_{100 \mu m} / L_{opt}$, versus  $L_{100 \mu m} / L_{radio}$.
Filled symbols correspond to the radio-loud population, while open
ones to the radio-weak population.

{\bf Figure 8} A colour-colour plot showing the behaviour of the rest-frame
luminosity ratio, $L_{1mm} / L_{radio}$, versus  $L_{1mm} / L_{opt}$.
Symbols as in figure 7.
\end{document}